\documentclass[twocolumn,showpacs,nofootinbib,preprintnumbers,amsmath,amssymb,superscriptaddress]{revtex4}
\usepackage{graphicx}% Include figure files
\usepackage{dcolumn}% Align table columns on decimal point
\usepackage{bm}% bold math
\def\lsim{\mathrel{\raise.3ex\hbox{$<$\kern-.75em\lower1ex\hbox{$qf$}}}}
\def\gsim{\mathrel{\raise.3ex\hbox{$>$\kern-.75em\lower1ex\hbox{$\sim$}}}}

\begin{document}
%put this where you want the line numbers to start

%\preprint{APS/123-QED}

\title{Improved Limits on Spin-Dependent WIMP-Proton Interactions from a Two Liter CF$_3$I Bubble Chamber}

\author{E. Behnke}
\affiliation{Indiana University South Bend, South Bend, USA}
\author{J. Behnke}
\affiliation{Indiana University South Bend, South Bend, USA}
\author{S.J. Brice}
\affiliation{Fermi National Accelerator Laboratory, Batavia, USA}
\author{D. Broemmelsiek}
\affiliation{Fermi National Accelerator Laboratory, Batavia, USA}
\author{J.I. Collar}
\affiliation{Enrico Fermi Institute, KICP and Department of Physics,
University of Chicago, Chicago, USA}
\author{P.S. Cooper}
\affiliation{Fermi National Accelerator Laboratory, Batavia, USA}
\author{M. Crisler}
\affiliation{Fermi National Accelerator Laboratory, Batavia, USA}
\author{C.E. Dahl}
\email{cdahl@kicp.uchicago.edu}
\affiliation{Enrico Fermi Institute, KICP and Department of Physics,
University of Chicago, Chicago, USA}
\author{D. Fustin}
\affiliation{Enrico Fermi Institute, KICP and Department of Physics,
University of Chicago, Chicago, USA}
\author{J. Hall}
\email{jeter@fnal.gov}
\affiliation{Fermi National Accelerator Laboratory, Batavia, USA}
%\affiliation{Fermi Center for Particle Astrophysics, Fermi National Accelerator Laboratory, Batavia, USA}
\author{J.H. Hinnefeld}
\affiliation{Indiana University South Bend, South Bend, USA}
\author{M. Hu}
\affiliation{Fermi National Accelerator Laboratory, Batavia, USA}
\author{I. Levine}
\affiliation{Indiana University South Bend, South Bend, USA}
\author{E. Ramberg}
\affiliation{Fermi National Accelerator Laboratory, Batavia, USA}
\author{T. Shepherd}
\affiliation{Indiana University South Bend, South Bend, USA}
\author{A. Sonnenschein}
\affiliation{Fermi National Accelerator Laboratory, Batavia, USA}
\author{M. Szydagis}
\affiliation{Enrico Fermi Institute, KICP and Department of Physics,
University of Chicago, Chicago, USA}

\collaboration{COUPP Collaboration}
\noaffiliation

\date{\today}% It is always \today, today,

%\linenumbers

\begin{abstract}
Data from the operation of a bubble
chamber filled with $3.5$ kg of CF$_{3}$I in a shallow underground site are reported.  
An analysis of ultrasound signals accompanying bubble nucleations
confirms that alpha decays generate a significantly louder acoustic 
emission than single nuclear recoils, leading to an efficient 
background discrimination.
Three dark matter candidate events were observed during an effective 
exposure of $28.1$ kg-day, consistent with a neutron background. This observation provides
strong direct detection constraints on WIMP-proton spin-dependent 
scattering 
for WIMP masses $>20$ GeV/c$^{2}$.
\end{abstract}

\pacs{29.40.-n, 95.35.+d, 95.30.Cq,   FERMILAB-PUB-10-318-A-CD-E}% PACS, the Physics and Astronomy
\maketitle

%\section{\label{introduction}Introduction}

There is abundant evidence that $\sim$$85\%$ of the matter in the Universe is cold, dark, and non-baryonic~\cite{dmevidence}.
Together, dark matter and dark energy are pillars in modern cosmology which form a simple framework to understand
 the detailed observations of baryonic structure, the Hubble diagram, the cosmic microwave background, and the abundances of
light elements.  However, present knowledge of the nature of these essential ingredients of cosmology is limited.  A new weak-scale
symmetry, such as the preservation of R-parity in supersymmetry, would predict a massive particle with properties that conform with the current
knowledge of dark matter~\cite{wimptheory}.  If these Weakly 
Interacting Massive Particles (WIMPs) are the
dark matter, then they may scatter off nuclei with enough energy, and at a high enough rate, to be detectable in the laboratory~\cite{wimpdetection}.

The Chicagoland Observatory for Underground Particle Physics (COUPP) 
employs the bubble chamber technique
to search for WIMP-nucleon elastic scattering \cite{COUPPtechnique}.  The bubble chamber is a powerful device to explore
nuclear recoils in a rare event search.  If the chamber pressure and temperature are chosen appropriately, electron recoils from
the abundant gamma-ray and beta-decay backgrounds simply do not nucleate bubbles~\cite{COUPPscience}.   
Neutrons from spontaneous fission and (alpha,n) in materials at the experimental site can be moderated with
 low-Z shielding materials, and cosmogenic neutrons can be reduced to negligible levels with an appropriate overburden. 
 Due to the threshold nature of the bubble nucleation process, alpha-decays are more of a background concern 
 for dark matter searches with superheated liquids than they are for technologies with event-by-event energy measurements.  
 Recently, however, the PICASSO collaboration reported event-by-event alpha-decay identification based on the acoustic emission
from bubbles \cite{PICASSOdiscrimination}.  This letter confirms the observation of alpha discrimination 
and utilizes this technique to set limits on spin-dependent WIMP-proton scattering.

This letter reports results from a $3.5$ kg CF$_3$I bubble chamber operated from August $19$th to December $18$th, 2009, in the MINOS near detector tunnel \cite{numi} at 
the Fermi National Accelerator Laboratory.  
The bubble chamber was located 3 meters off the axis of the NuMI 
neutrino beam~\cite{numi}, which operated with $10$ $\mu$s pulses, typically every 2.5 seconds.   
The short duty factor allowed for a calibration source of efficiently tagged, beam-induced, fast neutrons.  
Additional calibration was provided using a switchable americium-beryllium neutron source (sAmBe)~\cite{COUPPscience}.

The 300 ft overburden of the NuMI site was sufficient to shield out the hadronic component of the cosmic ray flux and to significantly attenuate the cosmic ray muon flux.   Residual cosmic ray muons passing near the bubble chamber were tagged using a 1000 gallon Bicron 517L liquid scintillator counter equipped with $19$ RCA-2425 photomultiplier tubes which surrounded the experiment on the sides and above.  The liquid scintillator, in conjunction with passive high-density polyethylene shielding below, provided a moderator for neutrons originating from natural radioactivity in the cavern rocks.

The bubble chamber consisted of a $150$ mm diameter $3$-liter synthetic fused silica \cite{suprasil} bell jar sealed to a flexible stainless steel bellows and immersed in propylene glycol within a stainless steel pressure vessel.  The propylene glycol, which served as the hydraulic fluid to manage the inner pressure of the bubble chamber, was driven by an external pressure control unit.  The flexible bellows served to ensure that the contents of the bell jar were at the same pressure as the hydraulic fluid.  The bell jar contained $3.5$ kg of CF$_3$I topped with water.  The water provided a buffer to ensure that the CF$_3$I contacted only the smooth silica and not the rough stainless steel surfaces.

The thermodynamic conditions of the chamber were monitored with two 
temperature sensors mounted on the bellows flanges and with three 
pressure transducers.  Four lead zirconate piezoelectric acoustic 
transducers were epoxied to the exterior of the bell jar to 
record the acoustic emissions from bubble nucleations, the audible ``plink'' used to trigger the flash lamps in early bubble chambers~\cite{GlaserPlink}.
Two VGA resolution CMOS cameras were used to photograph the 
chamber with a $20$-degree stereo angle at a rate of $100$ frames per second.  
Frame-to-frame differences in the image data provided the primary trigger for the experiment, 
typically initiating a compression within $20$ ms of a bubble nucleation.  
Stereo image data from the cameras were used to reconstruct the spatial coordinates of each bubble within the chamber.

The chamber was operated at a temperature of $29.5^\circ$C and a 
pressure of $26.5$ psia when sensitive to particle interactions, 
providing a Seitz model bubble nucleation 
threshold of 21 keV nuclear recoil energy~\cite{seitztheory}.  Although this threshold depends on a theoretical model, it has been benchmarked by comparing to the threshold for sensitivity to alpha decays~\cite{COUPPscience}.   Reasonable uncertainties in the measurements of the thermodynamic state of the superheated target, about $1$ psi and $0.5^\circ$C, correspond to a $15\%$ shift in the threshold of the chamber and a $30\%$ ($15\%$) change in sensitivity for $50$ ($100$) GeV WIMPs.

Following the nucleation of 
a bubble, the chamber was compressed to $215$ psia for $30$ seconds ($300$ seconds every $10$th event) 
to re-condense the CF$_3$I vapor.  During the compression period, data from the bubble event were logged, 
and the chamber was prepared for the next expansion.  To allow for equilibration of the bubble
chamber fluid, the valid live time for the chamber began $30$ seconds after an expansion.  From the outset, 
the performance of this chamber was a significant improvement over previous COUPP bubble chambers~\cite{COUPPscience}.  
There was no evidence of radon beyond a few dozen events immediately after the chamber fill.  
Vessel wall nucleations due to alpha decays, observed at a rate of $0.7$ events cm$^{-2}$ day$^{-1}$ in 
natural quartz vessels, were at an unmeasurably low level in this synthetic fused silica vessel.

% \cite{suprasil}.  In previous COUPP bubble chambers, the dominant sources of bubble nucleation were alpha emission from the walls of the natural quartz vessel and steady radon emanation in the chamber~\cite{COUPPscience}.  Cleanliness and removal of known radon emanators reduced the radon in the run reported here to an initial injection of a few dozen observed events decaying to a negligible rate.   Events near the walls are obvious in the steroscopic images, so are not a background, but the rate of nucleations on the walls does cause dead time.  The chamber was compressed for $60$ seconds each cycle and the first $30$ seconds of an expansion while the chamber stabilized to the operating pressure was not considered in the analysis.  Therefore, $1000$ cycles a day yields $\sim13\%$ live time in this detector.  The synthetic fused silica jar exhibited a rate of nucleations near the walls of the vessel of $\lsim 0.01$ events $/$ cm$^2$ $/$ day ($2$ per day in this chamber), at least two orders of magnitude less than the rate of $\sim 1$ event $/$ cm$^2$ $/$ day from the previously used natural quartz jar.   The rate of events from this synthetic fused silica vessel scaled to a one ton bubble chamber would result in a wall rate of $\sim 80$ per day, an acceptable loss of $\sim10\%$ dead time.

The data presented in this letter confirm PICASSO's observation of a difference
between the acoustic emissions from bubbles nucleated by nuclear
recoils and bubbles nucleated by alpha decays~\cite{PICASSOdiscrimination}. 
The current understanding of the physics behind this discrimination is that the ultrasonic acoustic emission from a growing bubble peaks
when the bubble is larger than a nuclear recoil track, tens of nanometers, but smaller than an alpha track, tens of microns\cite{russian}.  At the time of peak acoustic emission, a nuclear recoil track is a single bubble, while an alpha track may contain several microscopic bubbles making the alpha event louder.
%that the range of keV nuclear 
%recoils, like those expected from neutron or WIMP interactions, is tens of 
%nanometers, comparable to the critical radius of 
%formation of a single proto-bubble, the initial vapor volume of a critical bubble.  Alphas from radioactive decay are expected to create a larger 
%number of proto-bubbles 
%along their $\sim$$40$ micron path in addition to the proto-bubble created by the recoiling daughter nucleus.  These individual early 
%proto-bubbles are 
%expected to provide the dominant contribution to the ultrasound component of the 
%acoustic emission \cite{russian}. 
Because the COUPP bubble chamber fluid is a homogeneous medium, it is expected that acoustic alpha/nuclear recoil
discrimination should be more distinct than in the aqueous gel emulsion of superheated droplets used by PICASSO.

%Alpha decays in the bubble chamber are expected to produce more initial boiling sites, or proto-bubbles, than single nuclear recoils early in the event.  The heavy daughter nucleus will nucleate one proto-bubble, and the alpha particle will nucleate bubbles along its $\sim40$ micrometer track.  
%The resolution of the photography was about $0.5$ millimeters, too coarse to see the small-scale difference between alpha-decays and nuclear recoils.
%Calculations indicate that the acoustic power in bubble formation is released before the merger of proto-bubbles forms a single macroscopic bubble.  The power in the acoustic signal is, therefore, expected to be higher for alpha decays than for single nuclear recoils from neutron or WIMP interactions.  This forms the theoretical basis of the claim for alpha/neutron discrimination \cite{PICASSOdiscrimination} and of the analysis of the data from the acoustic sensors
%in this chamber.

\begin{figure}
\includegraphics[width=250 pt]{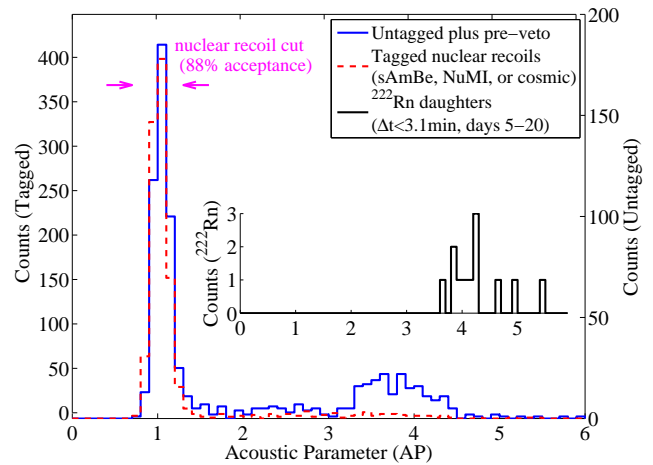}
\caption{\label{acoustic_discrimination} The acoustic parameter, defined in the text, 
is a measure of the loudness of the acoustic ``plink'' resulting from 
bubble nucleations.  The red, dashed histogram in the main panel is from the nuclear recoil calibration 
datasets (sAmBe, NuMI, cosmic).  The blue, solid histogram in the main panel is from all the untagged data, 
including data taken before the muon veto was commissioned.  The inset histogram contains events from $^{222}$Rn early in the run, 
which reproduce the second peak in the untagged dataset.  The 
acceptance of the nuclear recoil cut, indicated with arrows, is $88\%$.}
\end{figure}

The acoustic transducer signals were digitized with a $2.5$ MHz sampling rate and recorded for $100$ ms for each event.  
The sound of bubble nucleation showed a broad emission distinctly above background noise up to a frequency of $200$ kHz.
The analysis of these signals was based on an acoustic parameter ($AP$) which is a measure of loudness.  The $AP$ is a frequency weighted acoustic power density integral, corrected for sensor gain and bubble position\footnote{
$AP=\sum_jG_j\sum_nC_n(\vec{x})\sum_{f_{min}^n}^{f_{max}^n} f \times psd_{f}^j$.  $C_n(\vec{x})$ are corrections for the position dependence of the bubble acoustic emission.  $\vec{x}$ is the position of the bubble calculated from the stereoscopic images.  $G_j$ is the gain factor for acoustic transducer $j$.
$f$ is frequency, $n=1\ldots3$ indicates three
frequency bands (5-30 kHz, 30-80 kHz, or 80-120 kHz),  $j$ identifies the acoustic sensor, $f_{min}$ and $f_{max}$ are 
the boundaries of the frequency band, and $psd_f^j$ is the power 
spectral density for the bin with center frequency $f$ for sensor $j$.}.  The $AP$ was scaled to have a value of unity at the peak observed in its distribution for bubbles induced by known
neutron sources (the sAmBe source, the pulsed NuMI beam, or
cosmic muons).
The $AP$ distribution from these known neutron sources is shown in Fig.~\ref{acoustic_discrimination}.  Also shown is the distribution of untagged data, mostly the $\sim$$300$ kg-day exposure before 
the scintillator veto was installed, which showed a peak at $AP=1$ similar to the calibration 
data, with an additional component with $3\le AP \le5$.  The low $AP$ peak in the untagged data is caused by nuclear
recoils and the higher $AP$ peak is understood to be caused by alpha decays.

%The separation between the nuclear recoil and alpha-decay populations is more striking than the result from superheated emulsions due to the ability to spatially reconstruct the bubble position using the camera images which we use to remove the position dependence in the detected sound.   Additionally, we only consider bubbles where all energy deposition is in the bulk of the fluid by removing bubbles near the walls and buffer fluid interfaces from the analysis; in superheated emulsions, there is a finite probability of partitioning the energy deposition from an alpha particle between active and inactive detector elements.

All data have been subject to a fiducial volume cut requiring the stereoscopically reconstructed bubble positions to be
at least $5$ mm away from the wall.  The acceptance of this fiducial volume 
cut was
$75\%$ for nuclear recoils from the sAmBe and NuMI calibration samples.
The definition of $AP$ and of the $AP$ cut were specified before the veto was installed, and neither was tuned based on the WIMP search data sample.  
The nuclear recoil acceptance of the $AP$ cut was measured to be $88\%$ in the fiducial volume using 
the NuMI coincident and sAmBe calibration events.

%The acceptance of the cut on the sAmBe data is $90\%$ and the 
%acceptance on the NuMI coincident events is $85\%$.  

There was no viable method of injecting alpha-emitting isotopes into 
this bubble chamber.  On the contrary, efforts
were focused on removing these from the active volume, achieving a rate of
$\simeq 0.7$ alpha decays per kg day. 
A small amount of $^{222}$Rn injected during the fill of the chamber 
was identifiable via its characteristic decay time.  
These events were selected by requiring a time difference between bubbles of $<3$ minutes, 
corresponding to the sequence of alpha decays from $^{222}$Rn 
decaying to $^{218}$Po (T$_{\frac{1}{2}}=3.1$ min) which 
then emits a second alpha particle.  This small sample of $10$ known alpha decay events shown in the inset of Fig.~\ref{acoustic_discrimination} is consistent with the peak observed in the untagged data
at $3\le AP \le5$, confirming the expectation of a 
louder alpha acoustic emission.

\begin{figure}
\includegraphics[width=250 pt]{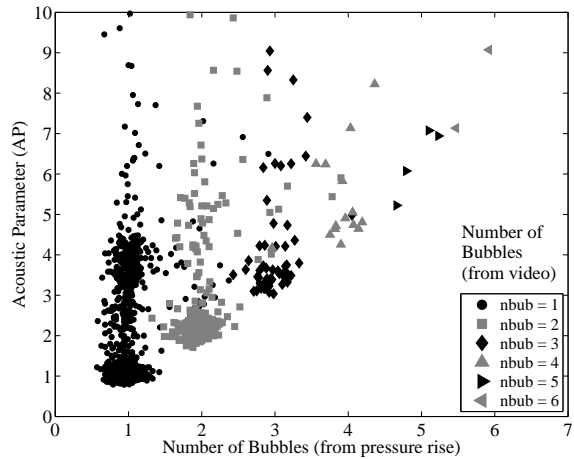}
\caption{\label{bubble_counting} 
The two methods of measuring the bubble multiplicity are plotted for 
the full data set of $291$ kg-day exposure.  The first method, based on automated video 
reconstruction software, searches for clusters of pixels changing between frames and is represented 
by the various symbols labeled in the legend.  The second method of counting bubbles was to measure 
the dP/dt associated with an event.    The abscissa is the pressure rise, scaled to the number of bubbles.  The ordinate
is the acoustic parameter $AP$ (see text).  }
\end{figure}
\begin{figure}
\includegraphics[width=250 pt]{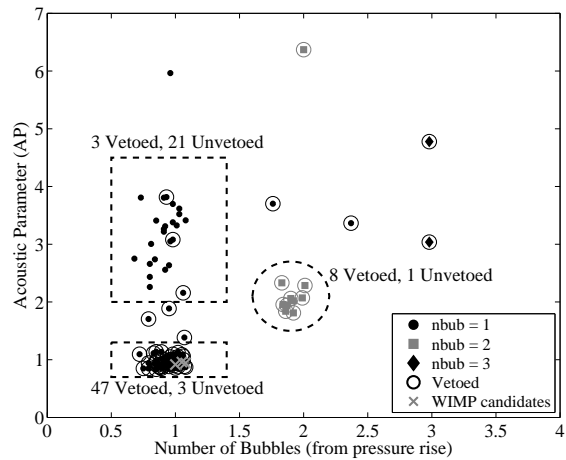}
\caption{\label{bubble_counts_wimp} 
Data from a $28.1$ kg-day WIMP search, plotted as in Fig.~\ref{bubble_counting}.  
In the alpha region (large dashed box) $21$ events with no evidence 
for muon activity were observed, with 3 events having coincident muon activity 
(labeled vetoed).  The smaller dashed box indicates 
the WIMP signal region, containing $3$ candidate events and $47$ events coincident with 
muons in the veto. The neutron double scatter region, indicated with a 
dashed circle, contains $8$ vetoed events and $1$ unvetoed event, 
the latter indicating an irreducible neutron background contamination (see text).}
\end{figure}
\begin{figure}
\includegraphics[width=250 pt]{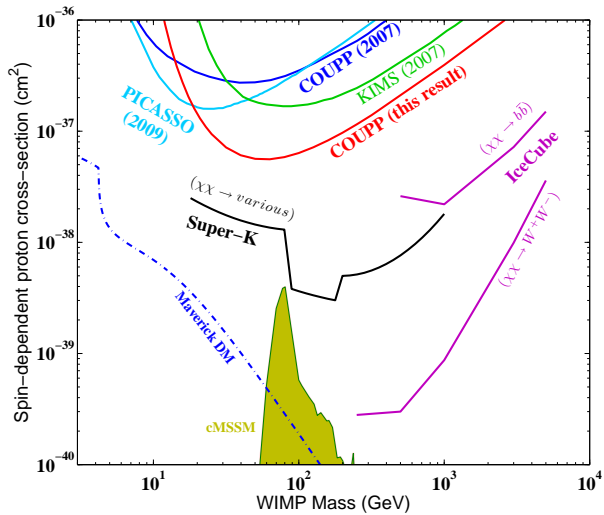}
\caption{\label{SDlimit} 
Improved limits on spin-dependent WIMP-proton elastic scattering from 
the data 
presented in this letter.  A previous COUPP result~\cite{COUPPscience} is shown for comparison.  
Direct detection limits from the PICASSO experiment~\cite{PICASSOlimit}, cyan, and the KIMS experiment~\cite{KIMSlimit}, green, 
are shown.  Limits on neutralino annihilation in the Sun from the IceCube~\cite{ICECUBElimit}, magenta,
and Super Kamiokande~\cite{SKlimit}, black, neutrino observatories are also plotted.  The indirect detection limits from the neutrino observations have additional dependence on the branching fractions of the annihilation products.  The gold region
indicates favored regions in cMSSM~\cite{cMSSMregion}. The blue, dashed-dotted line is the expected
cross-section for ``maverick'' dark matter with $\Omega h^2=0.1$~\cite{maverickdm}. }
\end{figure}

%The $AP$ is a measure of the number of
%proto-bubbles before they have merged to form macroscopic bubbles.  
The camera images and the slow pressure rise associated with bubble expansion provided two 
methods to count macroscopic bubbles in an event.  These bubble counting methods were used 
to produce Fig.~\ref{bubble_counting}, where the $AP$ is plotted as a function 
of bubble number inferred from the pressure rise.  The symbols 
indicate how many bubbles were inferred from camera images.    
WIMP interactions will nucleate only one bubble.  Neutrons are the only significant 
background able to produce
multiple macroscopic bubbles during an event. 
The $AP$ scales with the number of macroscopic bubbles.  
Tails in the $AP$ distribution for a given number of macroscopic 
bubbles are all towards 
louder values.\footnote{The high $AP$ tails are a possible indication of inefficiency for the $AP$ neutron selection, but not a background concern.  The nature of the events in the tail of the acoustic parameter has not been studied.  The rate is consistent with the expected rate of (n,alpha) reactions on flourine and carbon.}  This is consistent with the $AP$ scaling linearly with 
acoustic energy, with a minimum acoustic energy produced
by single bubbles.

Previous COUPP calibration data taken at $30^\circ$C showed that neutron induced nuclear recoils with energy above the Seitz threshold had a $50\%$ chance of nucleating bubbles~\cite{COUPPscience}.  A comparison of MCNP simulations of single and multiple rates with the rates observed during sAmBe calibration is consistent with this $50\%$ efficiency for the data presented here.  The efficiency is expected to reach $100\%$ at higher operating temperatures~\cite{COUPPscience}; unfortunately, a mechanical failure forced an early termination of the run, restricting data-taking to $30^\circ$C.

The WIMP search data spanned November 19th to December 18th.  
They are shown in Fig.~\ref{bubble_counts_wimp} along with the signal acceptance regions for various classes of event.  
There were $23$ days of 
livetime with the muon veto active.  A number of data quality
cuts were imposed to remove periods of poor detector performance.  The most restrictive data quality cut was imposed to excise
intermittent noise on the acoustic signals, removing $42\%$ of the science run.  An 
effective exposure of $28.1$ kg-days remained after imposing all data quality 
cuts, the fiducial volume cut, and the $AP$ nuclear recoil cut.
In the signal acceptance region there were $3$ WIMP candidate events and $47$ events coincident with muon 
activity in the veto.  In the alpha region there were $24$ events, 
$3$ of which were coincident with muon 
activity. These are statistically consistent with the high-AP tail expected from the
$47$ cosmic ray events in the signal region.  Of $9$ events with two macroscopic bubbles, 
one was not coincident with the veto.  This event is proof of an unvetoed 
neutron background component with sufficient rate to result in the 
observed 3 single-scatter WIMP candidate events.

Assessments of event-by-event discrimination against alpha decays and of the sensitivity to WIMP-nucleon scattering 
are limited by the unvetoed neutron background in this dataset.
Interpreting the three events in the signal region as alpha decays results in a conservative $90\%$ C.L. upper limit on 
the binomial probability of an alpha decay registering in the nuclear recoil 
signal region of $<26\%$.  
%by calculating 
%the mean binomial probability that results in three or fewer events 
%in the lower APr peak $10\%$ of the time, from a total of 24 trials.
At this operating pressure and temperature, an 
alpha particle will create bubble nucleation sites along its entire track, and there is clear evidence of a neutron background from the multiple scatter events, so these three events are likely not alpha decays. Therefore the 
presently derived alpha
background rejection should be considered a conservative 
assessment for the potential of this technique. We expect 
an improved estimate from runs in a deeper underground site, 
where the residual neutron background should be absent.

Interpreting the three events in the 
signal region as WIMP candidates results in a $90\%$ Poisson upper limit of $6.7$ for the mean of the signal.
The resulting improved limits on spin-dependent WIMP-proton couplings are shown in Fig.~\ref{SDlimit}.
The spin-independent sensitivity that can be 
extracted from present data is comparable to
that obtained by CDMS in another shallow underground facility~\cite{CDMSsuf}.  
The calculations assume the standard halo parameterization~\cite{lewinandsmith}, with $\rho_D = 0.3$ GeV c$^{-2}$ cm$^{-3}$, $v_{esc}=650$ km/s, $v_E=244$ km/s, $v_0=230$ km/s, and the spin-dependent couplings from the compilation in Tovey {\it et al.}\cite{spindependentcouplings}.
This result is consistent with a background from neutrons induced by residual cosmic 
radiation in the shallow site.

In view of the $\sim10^{-11}$ intrinsic 
rejection against minimum ionizing backgrounds~\cite{COUPPscience} 
and the acoustic alpha rejection demonstrated in this letter, 
a leading sensitivity to both spin-dependent and -independent 
WIMP couplings can be expected from the operation of CF$_3$I bubble 
chambers deep underground.
A first exploration of the cMSSM spin-dependent parameter space~\cite{cMSSMregion} of 
supersymmetric dark matter candidates is 
expected from operation of this chamber in a deeper site. 
At the time of this writing, a 60 kg 
CF$_3$I COUPP bubble chamber is being commissioned.

%\begin{acknowledgments}

The COUPP collaboration would like to thank Fermi National Accelerator Laboratory, 
the Department of Energy and the National Science Foundation for their support including grants 
PHY-0856273, PHY-0555472, PHY-0937500 and PHY-0919526.  We acknowledge technical assistance from 
Fermilab's Computing, Particle Physics, and Accelerator Divisions, and from E. Greiner, P. Marks, 
B. Sweeney, and A. Vollrath at IUSB.

%\end{acknowledgments}

\end{document}